%% file: sn-article.tex
\documentclass[sn-aps]{sn-jnl}
\usepackage{graphicx}%
\usepackage{multirow}%
\usepackage{amsmath,amssymb,amsfonts}%
\usepackage{amsthm}%
\usepackage{mathrsfs}%
\usepackage[title]{appendix}%
\usepackage{xcolor}%
\usepackage{textcomp}%
\usepackage{manyfoot}%
\usepackage{booktabs}%
\usepackage{algorithm}%
\usepackage{algorithmicx}%
\usepackage{algpseudocode}%
\usepackage{listings}%
\usepackage{subcaption}

\begin{document}

\title[Article Title]{\centering GPT-Invest{\fontsize{14}{16}\selectfont AR}: \\
Enhancing Stock Investment Strategies through Annual Report Analysis with Large Language Models}




\author{\fnm{Udit} \sur{Gupta}}





\input{sections/abstract.tex}




\maketitle

\input{sections/introduction.tex}

\input{sections/background.tex}

\input{sections/data.tex}

\input{sections/methods.tex}

\input{sections/results.tex}

\input{sections/conclusion.tex}

\bibliography{sn-bibliography}

\end{document}

%% file: sections/abstract.tex
\abstract{Annual Reports of publicly listed companies contain vital information about their financial health which can help assess the potential impact on Stock price of the firm. These reports are comprehensive in nature, going up to, and sometimes exceeding, 100 pages. Analysing these reports is cumbersome even for a single firm, let alone the whole universe of firms that exist. Over the years, financial experts have become proficient in extracting valuable information from these documents relatively quickly. However, this requires years of practice and experience. This paper aims to simplify the process of assessing Annual Reports of all the firms by leveraging the capabilities of Large Language Models (LLMs). The insights generated by the LLM are compiled in a Quant styled dataset and augmented by historical stock price data. A Machine Learning model is then trained with LLM outputs as features. The walkforward test results show promising outperformance wrt S\&P500 returns. This paper intends to provide a framework for future work in this direction. To facilitate this, the code has been released as open source\footnote{\href{https://github.com/UditGupta10/GPT-InvestAR}{Github Repository}}.\footnoteA{\hspace{0.25cm}Kindly contact to address any inquiries or provide feedback: \href{mailto:udit2008.gupta@gmail.com}{Email}, \href{https://www.linkedin.com/in/uditgupta10/}{LinkedIn}}}

%% file: sections/introduction.tex
\section{Introduction}\label{sec: Intro}
The stocks considered for this paper are from the US stock universe. Specifically, the top 1500 companies by market cap in the US which are part of the S\&P 500 largecap index, S\&P 400 midcap index and S\&P 600 smallcap index. Publicly listed companies in the US are required to file a comprehensive annual report to the U.S. Securities and Exchange Commission (SEC). These reports are called 10-K filings. They provide a detailed and standardized overview of a company's financial performance, operations, and potential risks, offering investors and regulatory bodies a comprehensive insight into the company's health and prospects. It acts as a critical source of information for investors, analysts, and other stakeholders to make informed decisions about investing in or interacting with a company. The 10-K filing of a company is used by investors to assess the company's financial statements, including the balance sheet, income statement, and cash flow statement. For the purpose of this paper, the terms 10-K and Annual report are used interchangeably.

In addition to providing company's financial statements, including the balance sheet, income statement, and cash flow statement, the 10-K also provides other valuable insights which are not directly captured by financial metrics and ratios. These include sections such as Business Overview, Risk Factors, Management's Discussion and Analysis (MD\&A), Legal Proceedings to name a few. Evaluating these sections is challenging because they cannot be expressed as a single numerical value. Their analysis is subjective, influenced by the reader's perspective, experience, and the context in which they are viewed.

Recently, Large Language Models (LLMs) such as GPT-3.5 (also known as ChatGPT) have emerged as powerful tools for enhancing the understanding and analysis of extensive documents including tasks like document summarization \cite{llm_documents}. In the context of Financial applications, A. Lopez-Lira et al. \cite{lopezlira2023chatgpt} show that LLMs can be successfully used for stock price prediction. For our use case, we explore the possibility of LLMs for answering complex questions that financial analysts might have regarding the company which can be addressed using the information from annual filings. An example of such a question could be \textit{``Does the company have a clear strategy for growth and innovation? Are there any recent strategic initiatives or partnerships?"}

%% file: sections/background.tex
\section{Background and Related Work}\label{sec: Background}
This paper utilises the GPT-3.5 version from OpenAI which is also the version used by ChatGPT currently. LLMs like GPT-3.5, having been trained on large corpus of data, have the capability of advanced Natural Language Understanding and can answer user queries based on the provided context. In their work, A. Lopez-Lira et al \cite{lopezlira2023chatgpt} construct a prompt containing a company specific news headline and ask the LLM to attribute the correct sentiment to that news headline for the purpose of determining stock price movement over the next day. The results obtained in their work showed that ChatGPT powered stock selection based on sentiment analysis exhibited a statistically significant predictive power. While their work offers valuable insights into the application of LLMs in finance, it also has certain challenges. First, they instruct the LLM to provide a ternary response for a prompt, which is used directly to make a buy or sell decision for the company's stock. In Machine Learning (ML) terminology, this method relies on a single feature (predictor variable) to forecast returns. This method works in practice as shown in their approach, but there may exist more complex interactions which could enhance performance. Second, because the stocks are chosen daily based on news headlines without taking into account portfolio turnover, this approach results in substantial transaction costs. In their paper, they demonstrate that including transaction costs has a notable impact on the strategy's net return. Specifically, when a per-transaction cost of 0.25\% is factored in, the net returns decrease dramatically, dropping from 500\% to 50\%. D. Blitz et al. \cite{blitz2023term} show that shorter time frame signals are not profitable after considering transaction costs and demonstrate that longer prediction horizons (six months, twelve months) can generate substantial positive net gains.

Taking inspiration from the work by A. Lopez-Lira et al \cite{lopezlira2023chatgpt} to apply LLMs in Investment decision making but also considering the downside of short term trading, this work introduces the notion of gathering insights from comprehensive company annual reports. As these reports are published annually, the signals generated from these are of longer duration and stock selection using these signals would incur negligible transaction cost per year. Moreover, in this paper, a Machine Learning model is used in conjunction with the output of the LLM to predict the best performing stocks over the upcoming year. The ML model is able to select relevant features while disregarding less significant ones. Additionally, it is capable of capturing intricate relationships among the features, ultimately leading to more accurate predictions.

%% file: sections/data.tex
\section{Data}\label{sec: Data} 
One of the main requirements of this exercise is fetching annual reports for top 1500 companies by market cap. SEC's Edgar database contains the historical 10-K filings by companies. These reports are fetched and stored locally along with their respective filing dates. 
For the purpose of this exercise the 10-K filings were retrieved spanning from the year 2002 to 2023, totaling 24,200 documents which occupy approximately 85GB of disk space.

The information in these reports help us answer specific questions regarding the company financials. When a question is asked pertaining to a company report, the relevant data in the 10-K acts as context information for the LLM to answer the question in the best possible way. Answers to these questions are used as input features to a Machine Learning model. More details regarding this are shared in Section \ref{subsec: LLM}.

In addition to generating features for the ML model, corresponding target values must also be assigned. Target value for a stock is derived from the percentage return of the stock between two successive annual report filings. To be conservative, the stock price two days after the filing is taken as starting point, and two days before the successive filing is taken as the ending point. In addition to capturing the return percentage from start to end, the return at intermediate intervals like 25th, 50th and 75th percentile of time is also captured. The maximum and minimum return achieved by the stock during this period is also computed. The details of deriving target values (from raw returns) for the ML model are discussed in Section \ref{subsec: targets}. In addition to computing the stock specific returns during the starting and ending period, the returns for the benchmark S\&P 500 index are also computed during the same period. This helps in comparing the stock portfolio returns with S\&P 500 returns when the ML model is applied on the test set.

The dataset is partitioned into training and testing sets, following the conventional practice in machine learning. This division facilitates model development, allowing for the assessment of model performance on unseen data. In this exercise, data from 2002 to 2017 is considered for training the ML model, and data from 2018 to 2023 is considered for testing. (\textbf{Note:} The annual reports published in 2023 are not considered but the historical price data is fetched until 2023 for evaluation purposes.)

An important factor to consider when developing applications using LLMs, such as the GPT-3.5 version by OpenAI, is the associated cost implications. Specifically, there is a cost associated with each question and answer measured in terms of the number of words (tokens) used and generated respectively. In addition to the question, sending the context from the 10-K report to the LLM also incurs cost. Taking into account these factors, 1k datapoints (out of 17.4k) were sampled from the training set for ML model building. Similarly, 500 datapoints (out of 6.8k) were sampled from the testing set for evaluation. From now on, in the paper, training and testing set refer to their respective sampled versions. The cost incurred for the complete exercise is around \$60. Apart from the monetary aspect, it also required significant processing time. The complete exercise of saving the embeddings of the documents and processing the questions using GPT-3.5 took approximately 50 hours for the sampled dataset. Expanding this exercise to the complete dataset would proportionately increase the time and cost involved.

%% file: sections/methods.tex
\section{Methods}\label{sec: Methods}
\subsection{Accessing Annual Reports}\label{subsec: AR}
To construct the dataset, access to historical 10-K filings of the top 1500 companies, ranked by market capitalization, is required. The list of these companies along with their stock symbols (tickers) are obtained from Wikipedia. While the filings are available in the SEC's archive database, the challenge lies in identifying the corresponding URLs. For this, a resource like \emph{Financial Modeling Prep} \cite{fmp} can be used to retrieve the URLs for all historical company filings. \emph{Note:} This website is gated via API access and may impose rate limits under the free access plan.

\subsection{Document Embeddings}\label{subsec: Embeddings}
For answering queries using the 10-K filing, it is essential to locate relevant sections within the document that can provide answers. These sections act as context for the LLM to formulate responses. Typically, the cosine similarity measure is used to assess the similarity between the query and document segments. This similarity calculation is based on vector representations of both the query and document chunks, often referred to as embeddings. These embeddings encode information into a fixed-size vector, capturing the document's semantic meaning in a high-dimensional space. Every word, sentence, and paragraph within the document contributes to shaping this vector. 

Several text embedding models are available, and their performance is evaluated through the \emph{Massive Text Embedding Benchmark (MTEB)} \cite{MTEB} to make an informed choice. Among these models, the \emph{all-mpnet-base-v2} \cite{mpnet} stands out with a strong MTEB score and efficient processing speed. Therefore, we select the \emph{all-mpnet-base-v2} model for text embedding in this exercise. For completeness, it's worth noting that OpenAI's embedding model, specifically \emph{text-embedding-ada-002} is also a good option. However, it was not used in this instance due to the substantial associated costs, especially considering the size of the documents. In contrast, the \emph{all-mpnet-base-v2} model \cite{mpnet} can be executed locally on a standard laptop within a reasonable timeframe, making it a more practical choice for this exercise.

Another subtle but important consideration involves the storage of these embeddings within a vector database. For this purpose, \emph{ChromaDB} \cite{chromadb} was selected for its seamless compatibility with \emph{LLama Index} \cite{Liu_LlamaIndex_2022} which is the primary LLM framework for this exercise.

\subsection{LLM for Feature Generation}\label{subsec: LLM}
The LLM used in this study is OpenAI's GPT-3.5-Turbo \cite{gpt}. Given the rapidly advancing nature of this field, the emergence of newer, more advanced models in the future is anticipated. The \emph{Llama Index} framework \cite{Liu_LlamaIndex_2022} streamlines the process of integrating LLM models from diverse sources. Furthermore, if an LLM model is open-source and its weights can be downloaded, it provides the flexibility to run the model either on a local machine or in a cloud environment. For this exercise, we have chosen to utilize GPT-3.5-Turbo \cite{gpt} as our LLM of preference due to its superior performance and the ease of running it via the OpenAI API.

For feature generation, specific questions are asked to the LLM regarding the financial and management aspects of the company. As previously detailed in Section \ref{subsec: Embeddings}, the embeddings relevant to the question are provided as context to the LLM. Utilizing the \emph{System Prompt}, the LLM is instructed to respond to the question by assigning a confidence score ranging from 0 to 100. This score is considered as a feature for the ML model. It is possible to generate multiple features by asking a variety of questions related to the company financial health. For example, one of the questions could be: \emph{''Does the company have a clear strategy for growth and innovation? Are there any recent strategic initiatives or partnerships?”}. Theoretically, many such questions can be posed and the resulting scores can serve as features for building a Machine Learning model. However, in practice, there is a cost associated with obtaining answers to additional questions. Although the questions themselves are concise, the context needed from the documents is relatively extensive, resulting in higher consumption of input tokens and subsequently an increase in cost. For this paper, a list of 27 questions was curated, the scores of which act as features for the downstream ML model building exercise. The process of crafting these questions is part of the domain known as Prompt Engineering.

\subsection{Label Creation}\label{subsec: targets}
For each stock ticker, the percentage return is calculated between two successive annual report filing dates. This gives the annual return of the stock denoted by \emph{target\_12m}. Similarly, percentage return is also calculated for the S\&P 500 index denoted by \emph{sp500\_12m}. Additionally, we determine the 98\textsuperscript{th} percentile (used as a proxy for the maximum) of return from the filing date represented as \emph{target\_max}. This value signifies the maximum return achieved by the stock during the annual period between two successive filings. Similarly, \emph{sp500\_max} is also computed for the S\&P 500 index.

The target value for the ML model is derived from \emph{target\_12m} and \emph{target\_max}. To achieve this, we refer Numerai's data documentation \cite{numerai}, which provides guidance on the methodology to construct the target variable using raw returns. The following steps are followed:
\begin{enumerate}
    \item The target values of stocks are assigned for each year separately. This is done to relatively rank returns of stocks within each year.
    \item The returns are first ranked, then normalised.
    \item The target values are confined to the range $[0,1]$. With 1 depicting higher returns.
    \item Lastly, the normalised returns are binned based on percentiles such that the target values are in range $[0,1]$.
\end{enumerate}

\subsection{Machine Learning Model}\label{subsec: ml model}
The dataset is split into training and testing sets. The data from 2002 to 2017 is considered for training the ML model, and data from 2018 to 2023 is considered for testing. The prediction task is formulated as a regression modeling problem where a continuous valued output is predicted. Following the prediction step, the buy strategy involves selecting the \emph{k} stocks with the highest predicted values within an year. 

The purpose of this exercise is to assess the predictive power of features generated by the LLM for selecting profitable stocks for the upcoming year. This is depicted using a straightforward yet robust machine learning model, specifically, Linear Regression with non-negative coefficients \cite{non_negative_linear_regression}. The choice of this configuration, which enforces non-negativity in the coefficients, was made to ensure that the relationships between features and the target variable are either positive or at the very least, non-negative. This alignment is crucial for our specific use case because the LLM-generated features were designed to exhibit a positive correlation with the target variable. Features that fail to predict the target variable in a consistently increasing manner will effectively receive zero coefficient values in this model. 

While more advanced techniques such as Gradient Boosted Decision Trees (GBDTs) can capture intricate feature relationships, they also require meticulous hyper-parameter tuning and regularization. The exploration of using GBDTs can certainly be undertaken as an independent exercise.

%% file: sections/results.tex
\section{Results}\label{sec: Results}
The analysis of predictions made on the test set reveals that the Machine Learning model constructed using GPT 3.5 features outperforms in selecting stocks, resulting in returns that surpass those of the S\&P 500 Index. The training and prediction task is performed within the Jupyter notebook titled \emph{modeling\_and\_return\_estimation.ipynb}.

\subsection{Finding the Ideal Number of Stocks to Buy}\label{subsec: num_ideal_stocks}
\begin{figure}[H]
    \centering
    \includegraphics[width=0.8\textwidth]{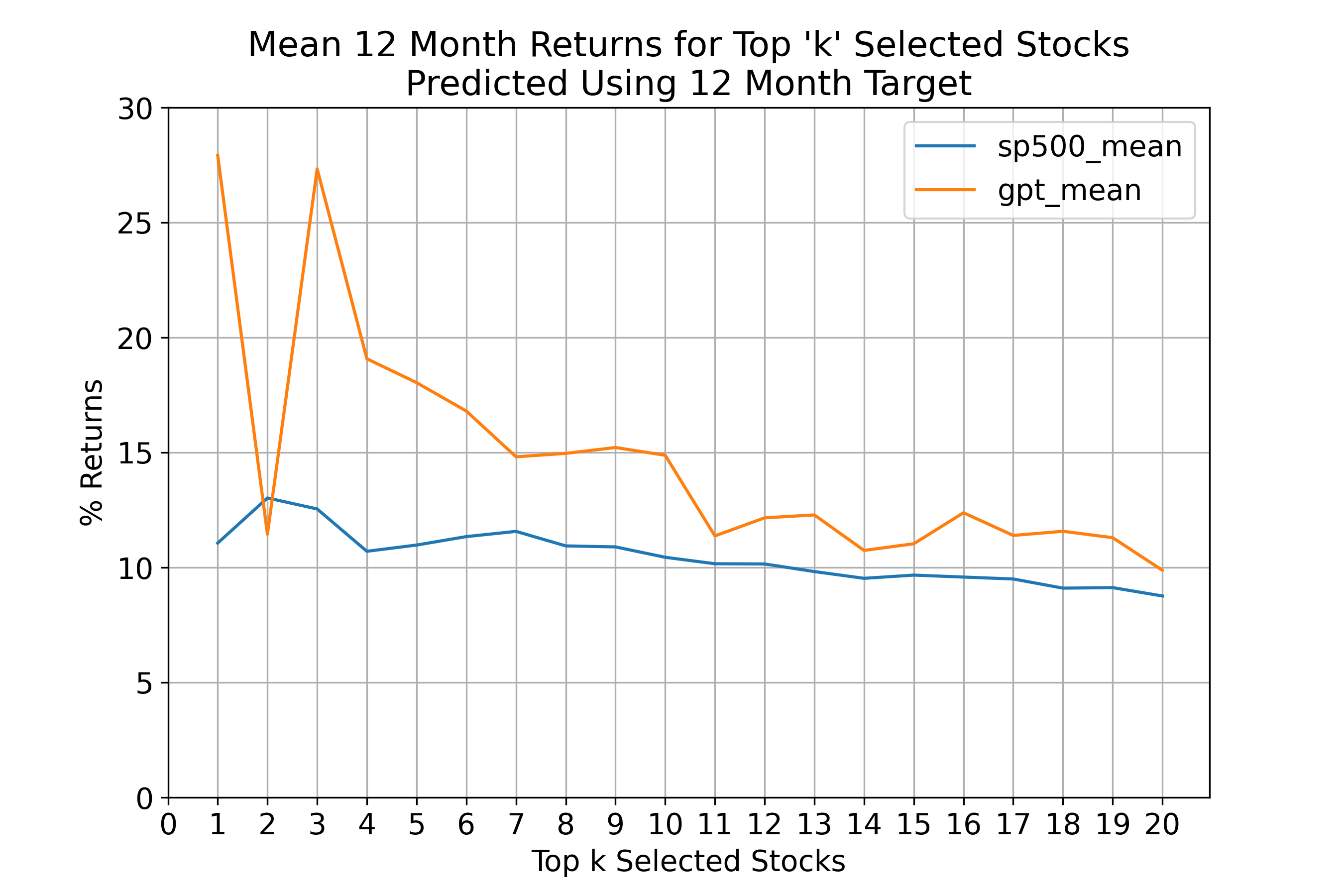}
    \caption{\centering
    Comparison of 12 Month Returns for various K values}
    \label{fig:12mReturns_12mTarget}
\end{figure}

Figure \ref{fig:12mReturns_12mTarget} compares the returns produced by the GPT model upon selecting \emph{top\_k} stocks vs the S\&P 500 returns for the same duration. The plot shows that the returns are higher for a smaller \emph{k} value and the returns diminish as \emph{k} value is increased. This shows that the higher ranked stocks (as predicted by the GPT model) provide better returns.\\
\textbf{Note:} The minor fluctuations in S\&P 500 returns for varying values of \emph{k} can be attributed to the fact that S\&P 500 returns are computed over the interval between successive annual report filings of individual stocks, and this duration varies among different stocks. This methodology ensures that both stock returns and S\&P 500 returns are measured over identical timeframes.

\begin{figure}[H]
    \centering
    \includegraphics[width=0.8\textwidth]{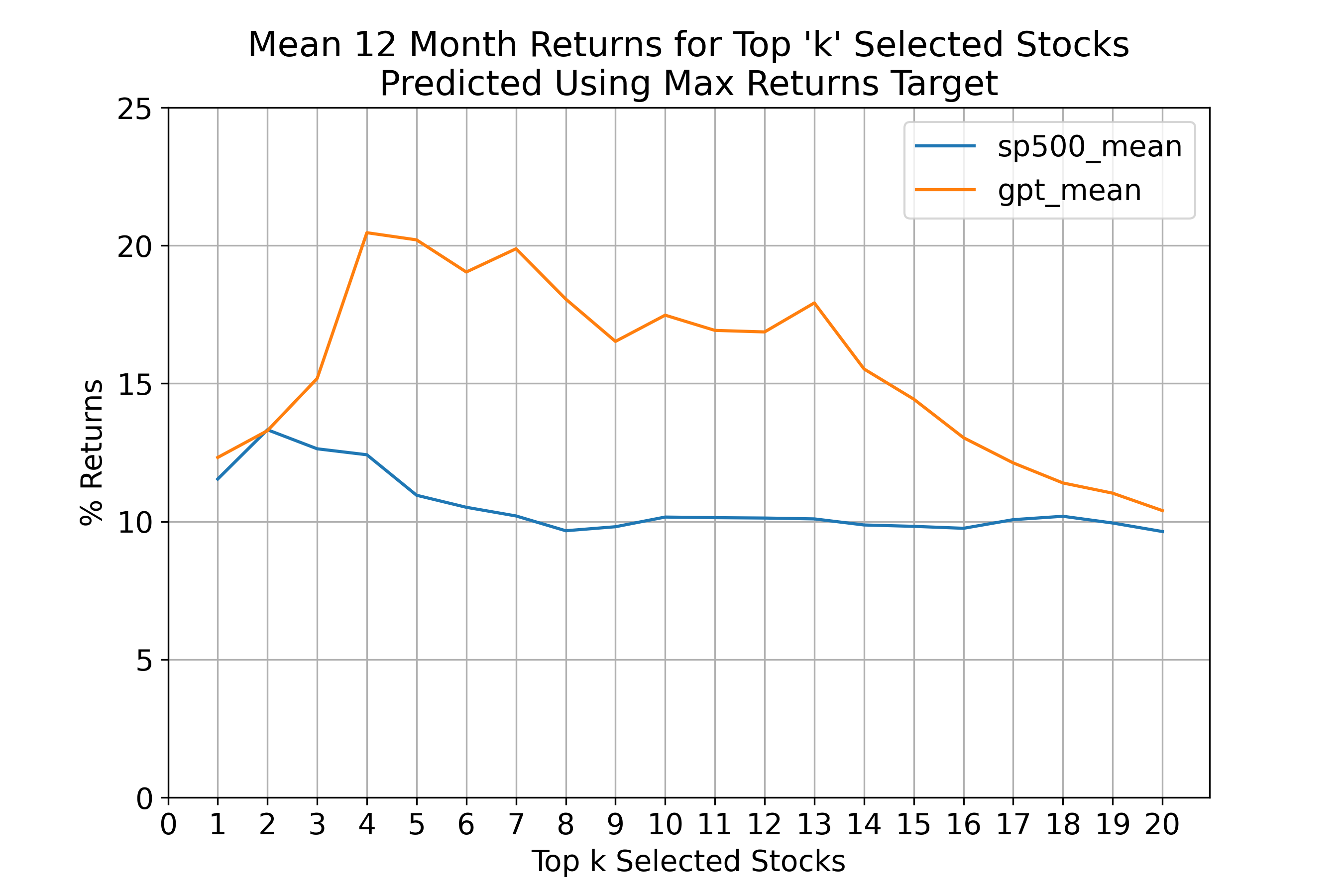}
    \caption{\centering
    Comparison of 12 Month Returns for various K values}
    \label{fig:12mReturns_maxTarget}
\end{figure}

Figure \ref{fig:12mReturns_maxTarget} also presents a comparison between the returns generated by GPT and the S\&P 500 returns. Both Figure \ref{fig:12mReturns_12mTarget} and Figure \ref{fig:12mReturns_maxTarget} display the mean returns over a 12-month period. The difference is that these figures are constructed using two distinct target variables. Figure \ref{fig:12mReturns_12mTarget} uses the 12 Month returns as the target variable, whereas Figure \ref{fig:12mReturns_maxTarget} uses the Max returns as the target variable.

A few notable observations can be made from the plots above. Firstly, for optimizing returns, it is advantageous to have a lower value of \emph{k}. In this context, a \emph{k} value of 5 appears to be a reasonable choice. When \emph{k} is set to 5, it signifies that 5\% of the available stocks are chosen for the buy strategy. To provide context, the test set comprises of 500 randomly sampled stocks over a 5-year period. Therefore, selecting 5 stocks per year equates to picking 5 out of 100 stocks annually, representing a 5\% selection rate.

\subsection{Analyzing Cumulative Returns for Different Strategies}\label{subsec: cum_returns}

\begin{figure}[H]
    \begin{subfigure}{0.5\textwidth}
        \centering
        \includegraphics[width=\linewidth]{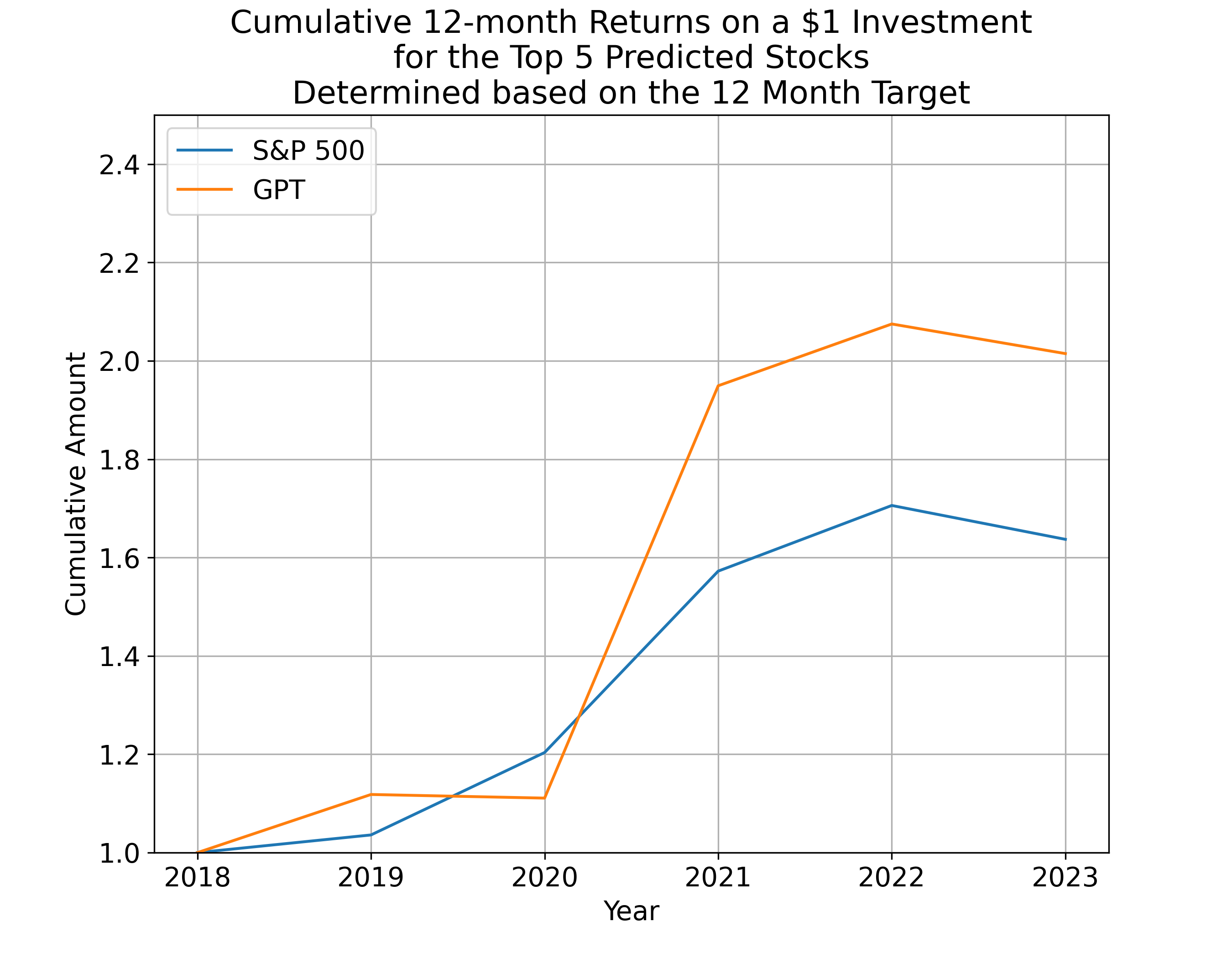}
        \caption{Modelled using 12-Month Returns Target}
        \label{fig:12mReturns_a}
    \end{subfigure}%
    \begin{subfigure}{0.5\textwidth}
        \centering
        \includegraphics[width=\linewidth]{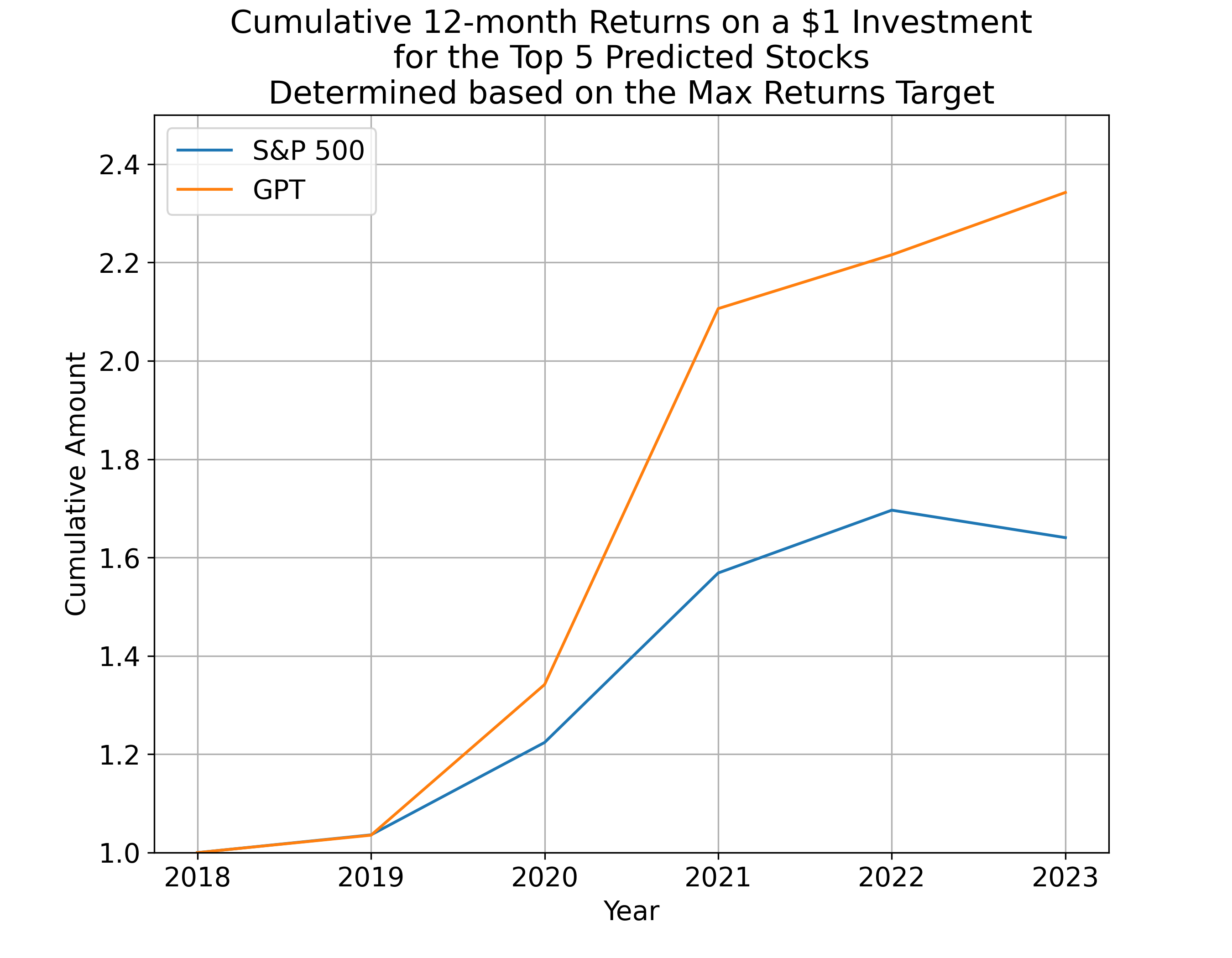}
        \caption{Modelled using Max Returns Target}
        \label{fig:12mReturns_b}
    \end{subfigure}
    \caption{\centering
    Cumulative 12-month Returns on a \$1 Investment for the Top 5 Predicted Stocks}
    \label{fig:12mReturns}
\end{figure}

Figure \ref{fig:12mReturns} depicts the investment performance of \$1 from the start of 2018 to the start of 2023 for two differently constructed models. While both models outperform the S\&P 500 returns during this period, the model illustrated in Figure \ref{fig:12mReturns_b} demonstrates significantly superior performance compared to that shown in Figure \ref{fig:12mReturns_a}. The key differentiator between the two approaches lies in the utilization of different target variables. This outcome underscores the importance of selecting an appropriate target variable for achieving better performance.

\begin{figure}[H]
    \begin{subfigure}{0.5\textwidth}
        \centering
        \includegraphics[width=\linewidth]{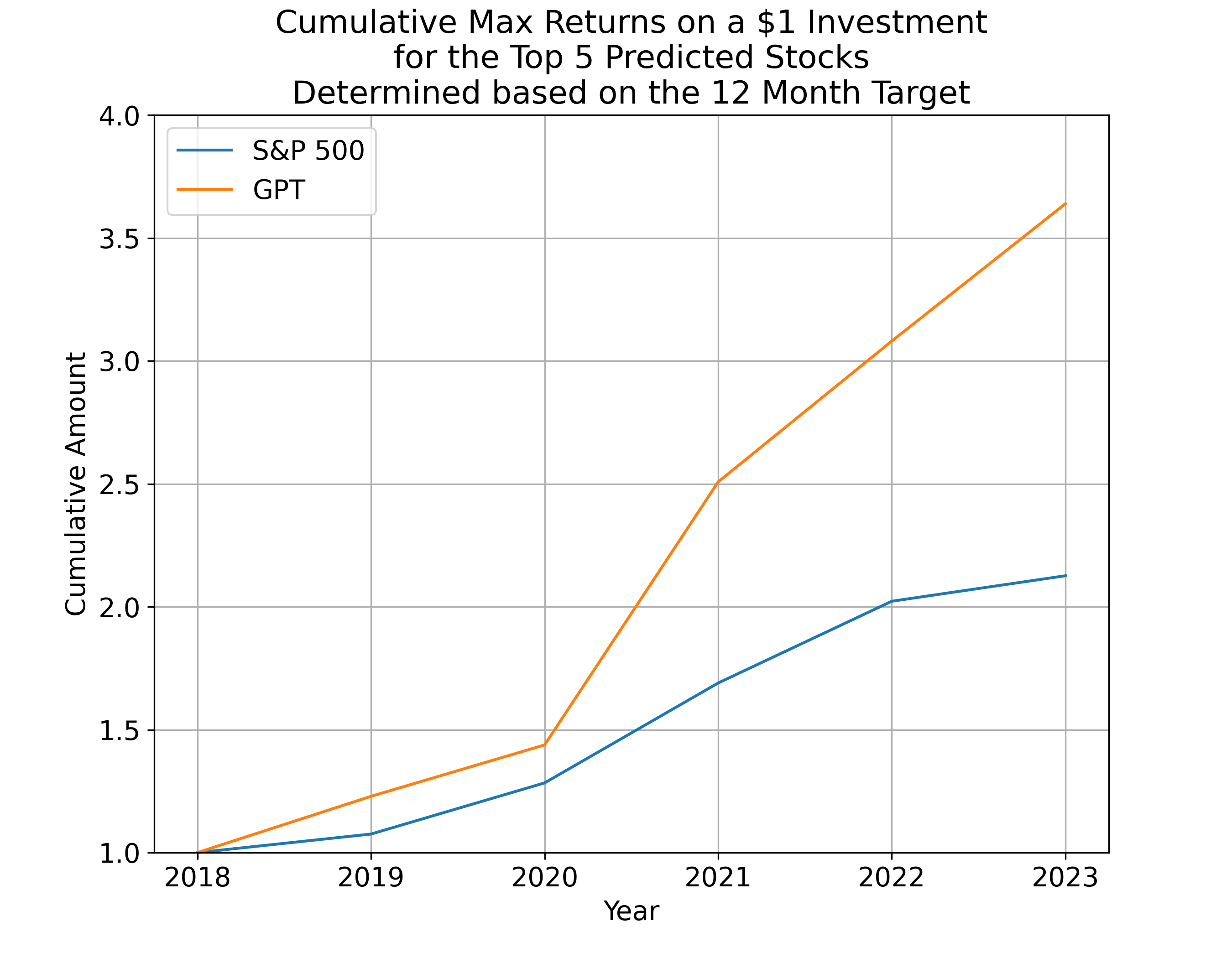}
        \caption{Modelled using 12-Month Returns Target}
        \label{fig:maxReturns_a}
    \end{subfigure}%
    \begin{subfigure}{0.5\textwidth}
        \centering
        \includegraphics[width=\linewidth]{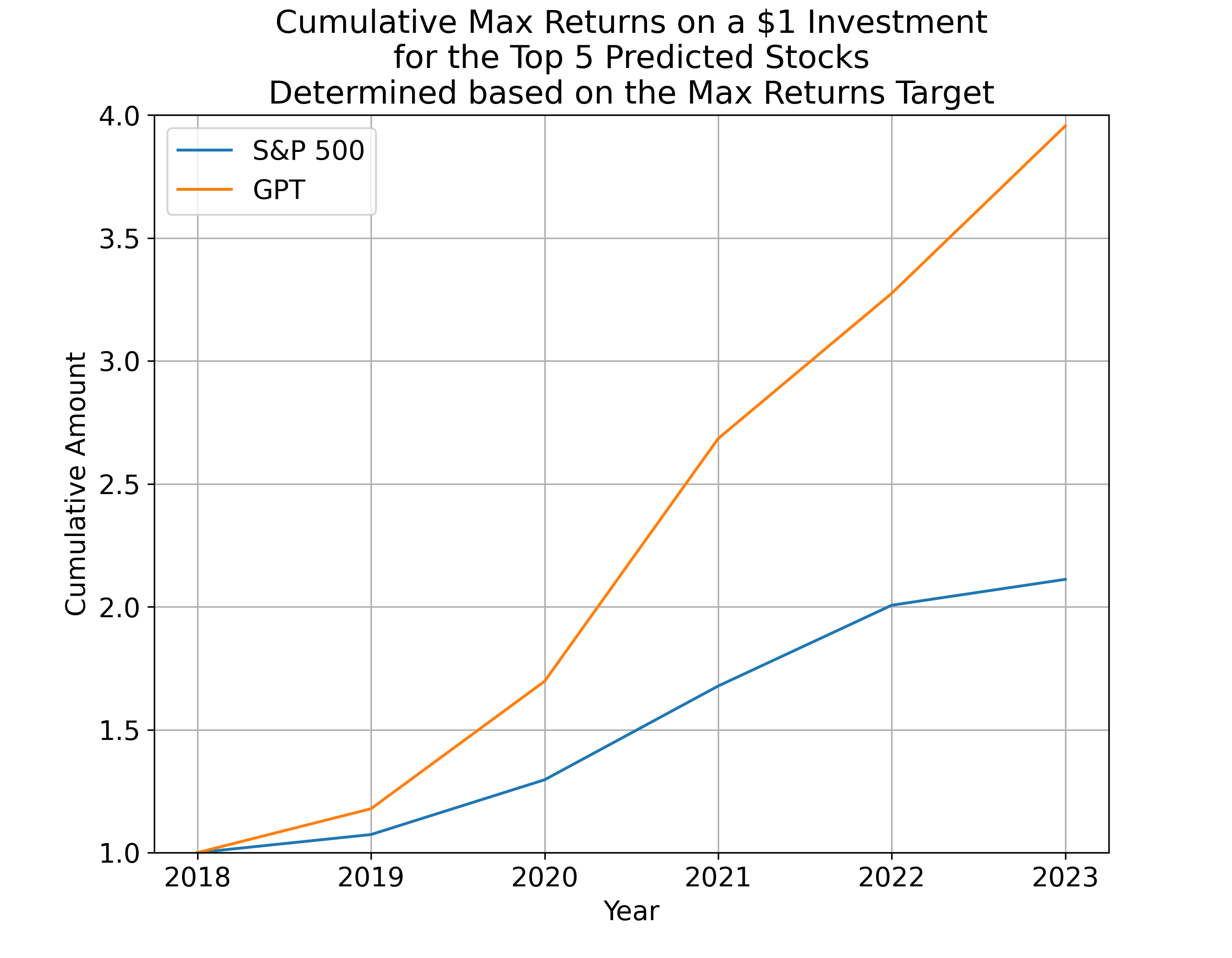}
        \caption{Modelled using Max Returns Target}
        \label{fig:maxReturns_b}
    \end{subfigure}
    \caption{\centering
    Cumulative Max Returns on a \$1 Investment for the Top 5 Predicted Stocks}
    \label{fig:maxReturns}
\end{figure}

Figure \ref{fig:maxReturns} presents an evaluation of the \emph{Max Returns} strategy. This strategy involves calculating returns from the start of the annual report date until the point representing the 98\textsuperscript{th} percentile of stock price within that year. In essence, it assumes that stocks are acquired shortly after the release of the annual report and sold at or near the highest price within that year. While this might appear to be an optimistic scenario, it's worth noting that the construction of S\&P 500 returns follows a similar methodology. Therefore, this analysis enables us to make a direct and equitable comparison of the \emph{Max Returns} strategy with the benchmark index.

Figure \ref{fig:maxReturns_b} shows the best case scenario where the \emph{Max Returns} strategy is able to quadruple the \$1 investment in 5 years whereas the S\&P would have only doubled the investment.

%% file: sections/conclusion.tex
\section{Conclusion}\label{sec: Conclusion}
As shown in the Results Section \ref{sec: Results}, the scores generated by an LLM, such as GPT-3.5 in this instance, can act as valuable features for constructing a Machine Learning model. There is flexibility in choosing and constructing the target (response) variable based on various timeframes. In this paper, two types of target variables have been studied and proven to be effective. Nevertheless, it is essential to recognize that the potential for alternative methods of defining the target variable remains open.

Depending on the choice of target variable, the ML model can outperform the benchmark returns of the S\&P 500 index. This is significant, because many of the existing actively managed techniques that employ short term trading strategies for generating alpha also incur significant transaction costs and hence may not be net profitable. The framework presented in this paper shows that longer time frame investing using LLMs can prove to be beneficial without the burden of excessive transaction costs.